\documentclass[aps,prl,preprint,superscriptaddress,showkeys,10pt]{revtex4-1}

\usepackage{graphicx}
\usepackage{epsfig}
\usepackage{amsmath,amsthm}
\usepackage{color}
\usepackage{verbatim}
\usepackage{ulem}

\newcommand{\red}{\color{black}}
\begin{document}

\title{
{ Direct Observation of Antiferromagnetic 
Parity Violation
 in the Electronic Structure 
of  Mn$_2$Au}} 

\author{O. Fedchenko}
\affiliation{Institut f\"{u}r Physik, Johannes Gutenberg-Universit\"{a}t Mainz, Staudingerweg 7, D-55099 Mainz, Germany}
\author{L. \v{S}mejkal}
\affiliation{Institut f\"{u}r Physik, Johannes Gutenberg-Universit\"{a}t Mainz, Staudingerweg 7, D-55099 Mainz, Germany}
\affiliation{Inst. of Physics Academy of Sciences of the Czech Republic, Cukrovarnick\'{a} 10,  Praha 6, Czech Republic}
\author{M. Kallmayer}
\affiliation{Surface Concept GmbH, Am S\"{a}gewerk 23A, D-55124 Mainz, Germany}
\author{Ya. Lytvynenko}
\affiliation{Institut f\"{u}r Physik, Johannes Gutenberg-Universit\"{a}t Mainz, Staudingerweg 7, D-55099 Mainz, Germany}
\author{K. Medjanik}
\affiliation{Institut f\"{u}r Physik, Johannes Gutenberg-Universit\"{a}t Mainz, Staudingerweg 7, D-55099 Mainz, Germany}
\author{S. Babenkov}
\affiliation{Institut f\"{u}r Physik, Johannes Gutenberg-Universit\"{a}t Mainz, Staudingerweg 7, D-55099 Mainz, Germany}
\author{D. Vasilyev}
\affiliation{Institut f\"{u}r Physik, Johannes Gutenberg-Universit\"{a}t Mainz, Staudingerweg 7, D-55099 Mainz, Germany}
\author{M. Kl{\"a}ui}
\affiliation{Institut f\"{u}r Physik, Johannes Gutenberg-Universit\"{a}t Mainz, Staudingerweg 7, D-55099 Mainz, Germany}
\author{J. Demsar}
\affiliation{Institut f\"{u}r Physik, Johannes Gutenberg-Universit\"{a}t Mainz, Staudingerweg 7, D-55099 Mainz, Germany}
\author{G. Sch{\"o}nhense}
\affiliation{Institut f\"{u}r Physik, Johannes Gutenberg-Universit\"{a}t Mainz, Staudingerweg 7, D-55099 Mainz, Germany}
\author{M. Jourdan}
\affiliation{Institut f\"{u}r Physik, Johannes Gutenberg-Universit\"{a}t Mainz, Staudingerweg 7, D-55099 Mainz, Germany}
\author{J. Sinova}
\affiliation{Institut f\"{u}r Physik, Johannes Gutenberg-Universit\"{a}t Mainz, Staudingerweg 7, D-55099 Mainz, Germany}
\affiliation{Inst. of Physics Academy of Sciences of the Czech Republic, Cukrovarnick\'{a} 10,  Praha 6, Czech Republic}
\author{H. J. Elmers}
\affiliation{Institut f\"{u}r Physik, Johannes Gutenberg-Universit\"{a}t Mainz, Staudingerweg 7, D-55099 Mainz, Germany}
\email{elmers@uni-mainz.de}

\begin{abstract}
Parity symmetric photoemission spectra 
are ubiquitous in solid state research, being 
prevalent in many highly active areas, such as unconventional superconductors, nonmagnetic and antiferromagnetic topological insulators, 
and weakly relativistic collinear magnets,
among others \cite{Sobota2021,Smejkal2021a}.
The direct observation of parity-violating\cite{Smejkal2017,Hayami2018,Watanabe2020,Elmers2020,Hayami2020} metallic Kramers degenerate bands has remained hitherto experimentally elusive. 
Here we observe the antiferromagnetic parity violation (APV) in the bandstructure of Mn$_{\text{2}}$Au thin films by using momentum microscopy with sub-$\mu$m spatial resolution, allowing momentum resolved photoemission on single antiferromagnetic domains. 
The APV arises from breaking the $\mathcal{P}$ symmetry of the underlying crystal structure  by the collinear antiferromagnetism, while preserving the joint space-time inverison $\mathcal{PT}$-symmetry and in combination with large spin-orbit coupling \cite{Smejkal2017,Elmers2020}. 
In addition, our work also demonstrates a novel
tool to directly image the N\'eel vector direction  by combining spatially resolved momentum
microscopy with ab-initio calculations.
\end{abstract} 


\maketitle

\newpage

{\red
Solid-state textbooks and angular-resolved photoemission spectroscopy (ARPES) studies commonly
 present 
 energy bands that are parity symmetric in momentum, i.e. $E(k)=E(-k)$.  
This is enforced  by the symmetries of the materials, such as inversion/parity $\mathcal{P}$,  time-reversal $\mathcal{T}$,   
time-reversal coupled  with 
 translation $\mathcal{T}\boldsymbol{t}$ 
 or time-reversal coupled with the spin rotational symmetry $\mathcal{TR}_{S}$, where $\mathcal{R}_S$ rotates the spin by 180$^\circ$. 
 {Even, the Rashba materials (and noncentrosymmetric spin-orbit coupled systems in general), which break parity only in  spin space, do preserve parity
 due to the $\mathcal{T}$ or $\mathcal{T}\boldsymbol{t}$ symmetry, i.e.  $E^{\uparrow}(k)=E^{\downarrow}(-k)$}.
 Therefore, from this perspective, bulk band structures violating parity in systems with Kramer degenerate bands are rather unique and rare.
Bands with broken parity have  appeared only in systems that break Kramers degeneracy by both spatial inversion and time symmetry at interfaces \cite{Carbone2016}, 
 and have been predicted to appear 
in complex noncoplanar magnets that break ${\cal T}R_S$ \cite{Hayami2018}. }

\begin{figure}[h]
\includegraphics[width=0.9\columnwidth]{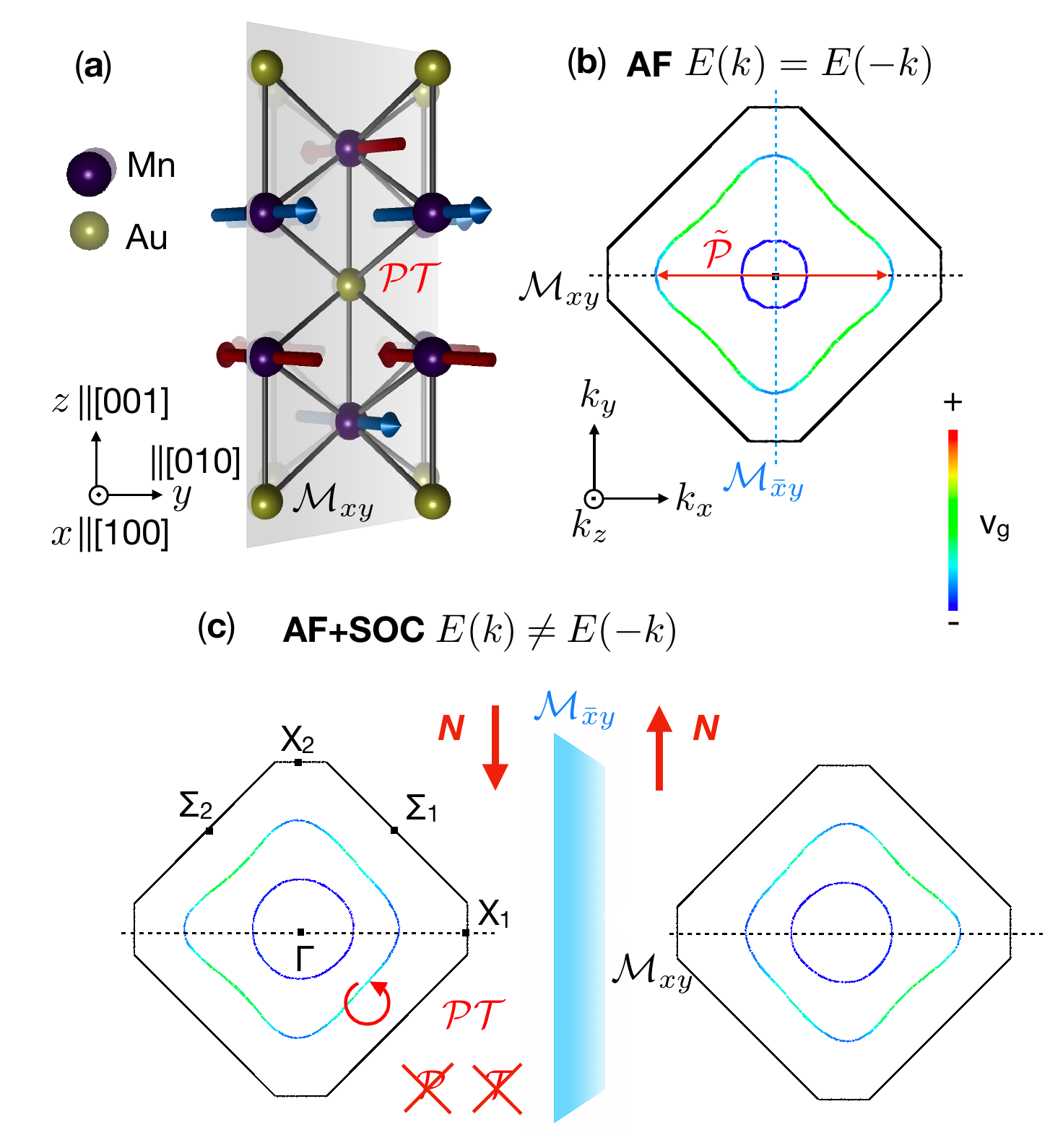}
\vspace{- 0.4 cm}
\caption{\label{FigConcept} 
(a) Structure of Mn$_2$Au 
demonstrating 
the ${\cal P T}$ symmetry in the magnetically ordered phase; 
(b) Calculated constant energy surface in momentum space at -0.4~eV  of Mn$_2$Au without SOC.
The  spin-group symmetry ${\cal R}_S{\cal T}$ in the decoupled spin and space sector preserve the bandstructure parity, 
acting as $\cal {P}$ in the band structure, denoted by $\tilde{\cal P}$. Color code denotes the group velocity $v_g$.  
(c) Asymmetric constant energy surface at energy -0.4 eV  arising from 
the N\'eel order  and SOC, 
shown for two directions of the Ne\'el order connected by the mirror symmetries ${\cal M}_{xy}$ and ${\cal M}_{\bar{x}y}$.
}
\vspace{- 0.5 cm}
\end{figure}

More recently,  APV,  
i.e.  $E(k) \neq E(-k)$, was predicted to arise in spin-orbit coupled 
collinear antiferromagnets such as CuMnAs and Mn$_{\text{2}}$Au, which break $\mathcal{P}$ and $\mathcal{T}$ symmetries but preserve its $\mathcal{PT}$ combination, as we show in Fig.~\ref{FigConcept}(a), in spite of having crystal parity, {\it i.e.}, the crystal is itself centrosymmetric but the antiferromagnetic order breaks that symmetry. 
Without spin-orbit coupling (SOC), the bandstructure is still symmetric [Fig.~\ref{FigConcept}(b)], {and its nonrelativistic symmetry group $P1_{4}/2_{m}1_{m}1_{m}$\cite{Smejkal2021a,Elmers2020} exhibits also the two mirror planes $\mathcal{M}_{xy}$ and $\mathcal{M}_{\bar{x}y}$ marked with dotted lines in black and blue color, respectively.}  
With SOC, the magnetic symmetry group  is $Fm'mm$\cite{Elmers2020} and breaks the $\mathcal{M}_{\bar{x}y}$ symmetry. 
The bandstructure cannot be superposed with its parity image related by the blue mirror plane $\mathcal{M}_{\bar{x}y}$, as we illustrate in 
Fig.~\ref{FigConcept}(c), exhibiting  directly APV.
Here $\cal{P T}$ maps a state to its equal momentum state (with opposite spin) since the presence of spin-orbit coupling prevents the coupling to states with opposite spin and momenta, hence preserving Kramers degeneracy in these systems.
Up to now, this predicted APV  was 
observed only indirectly through the electrical current induced manipulation of the N\'eel vector via N\'eel spin-orbit torques (NSOT) in CuMnAs \cite{Zelezny2014,Wadley2016} and Mn$_2$Au \cite{Bodnar2018}, and second-order magnetoresistance in CuMnAs \cite{Godinho2018}.

The direct observation of APV using conventional ARPES
is not possible due to averaging over many antiferromagnetic domains
with typical sizes in the $\mu$m range.
While 
a ferromagnetic capping layer can 
be used to orient the N{\'e}el vector via exchange coupling~\cite{Bommanaboyena2021},
the  capping layer prevents surface sensitive ARPES measurements.

In this Letter, we use instead time-of-flight momentum microscopy combined with sub-$\mu$m 
spatial resolution (sub-$\mu$-ToFMM) to directly observe the APV by an asymmetric photoemission intensity, $E(k) \neq E(-k)$.
Such a spatial resolution allows one to measure the electronic structure in areas restricted to single antiferromagnetic domains with well defined
N{\'e}el vector orientation.
We also further demonstrate below the direct imaging of the N\'eel vector direction, ${\bf N}$.



Epitaxial Mn$_2$Au(001) films \textcolor{black}{with a thickness of 40~nm} were grown \textcolor{black}{by rf-sputtering} on Al$_2$O$_3$(1$\rm{\bar{1}}$02) substrates 
with a Ta(001) buffer. 
Mn$_2$Au has a body centered tetragonal crystal structure (bct$_2$), 
with the (001) plane exhibiting a 4-fold structural symmetry and space group $I4/mmm$. 
Details of the sample growth, characterization by X-ray and electron diffraction, as well as by atomic force microscopy, are reported in Refs.~\cite{Jourdan2015,Satya2020}.
During growth, the magnetic in-plane anisotropy aligns 
the N{\'e}el vector nearly equally distributed along both 
magnetic $\langle 110 \rangle$ easy axes with domain sizes 
in the micrometer range~\cite{Sapozhnik2018}. 
Details of the sample, ARPES, ToFMM and photoemission measurements are given in the Supplementary Material.



In the experiment we first identify antiferromagnetic domains in real space
by magnetic linear dichroism (MLD) in the photoemission electron microscopy (PEEM) mode.
We then select single domain areas by the field aperture. Lastly, we record the photoemission intensity in momentum space
by setting the electron optics
 to momentum mode.

\begin{figure}
\includegraphics[width=\columnwidth]{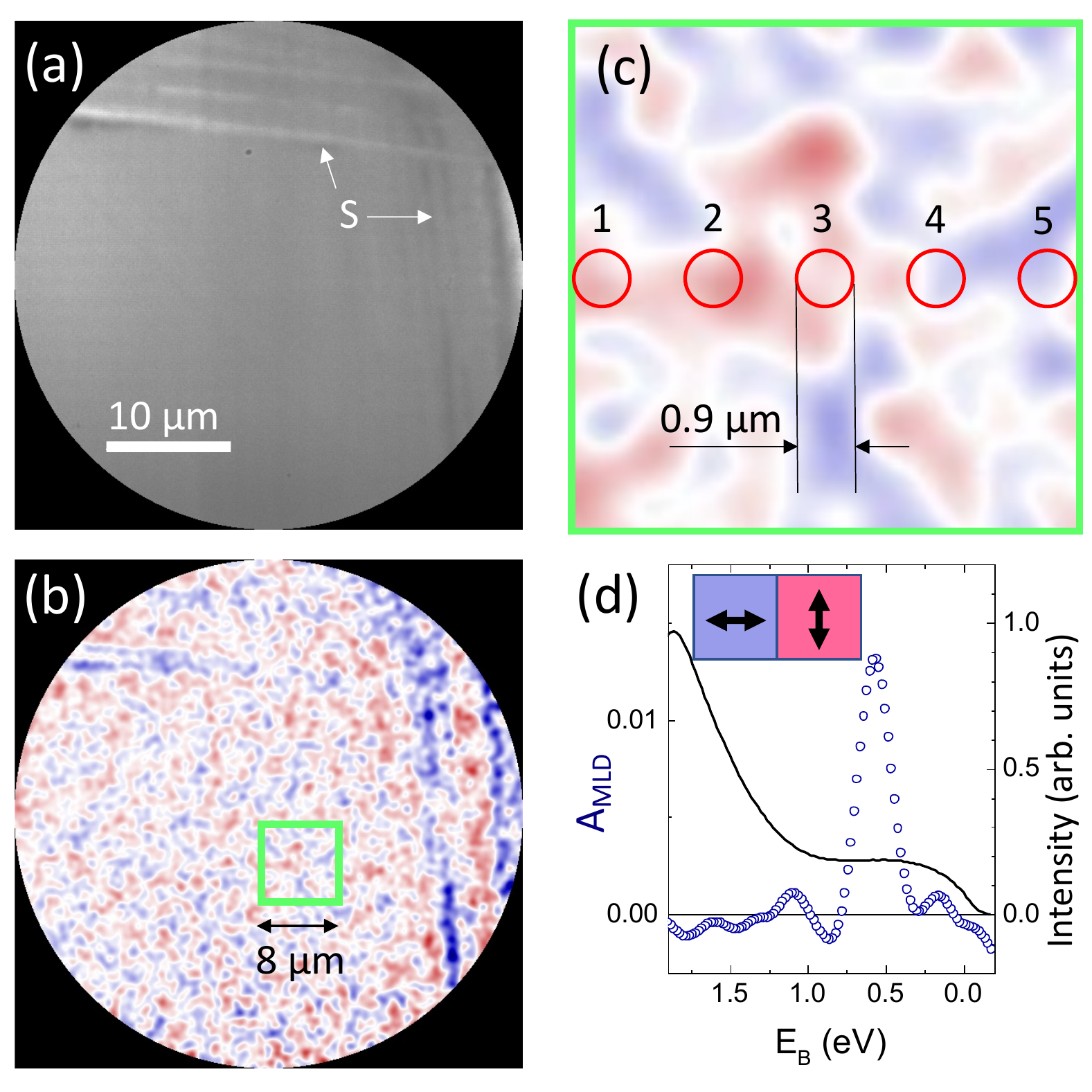}
\vspace{-0,6 cm}
\caption{\label{Fig1} 
(a) PEEM image of the Mn$_2$Au(100) sample surface obtained with 6.4~eV photon energy.
Scratches (S) on the otherwise homogeneous surface serve for position determination.
(b) Magnetic linear dichroism (MLD) image for the 
area as in (a) with color (red/blue) coded asymmetry $A_{\rm MLD}=(I_p-I_s)/(I_p+I_s)$ 
($E_B=0.6$~eV, $p$- and $s$-polarized light). 
(c) Magnified image 
from the green square indicated in (b).
The five numbered circles define the regions of interest 
selected by the field aperture that are used for momentum microscopy 
[results shown in Fig.~\ref{Fig2}(g-1 to g-5)].
(d) Intensity (black line) and MLD asymmetry, $A_{\rm MLD}$, (circles) vs. $E_B$. 
}
\vspace{-0,6 cm}
\end{figure}
The ToFMM with the electron optics set to the PEEM mode
detects the spatial distribution of photoemission intensity, where 
the sample surface is magnified on the detector with a field-of-view of 50~$\mu$m in diameter [Fig.~\ref{Fig1}(a)].
The spatial resolution in this mode is limited by the spherical aberration of the objective lens
that increases with increasing parallel momentum of the detected electrons.
The time-of-flight detection mode allows in addition to measure the kinetic energy of the photo-emitted electrons,
hence simultaneously acquiring the three-dimensional data array $I(E_B,x,y)$.

In order to extract the magnetic contrast from the images, we exploit MLD.
For reflected light, MLD originates from a magnetic-linear birefringence 
being sensitive to the magnetic order axis instead of its direction~\cite{Hubert1998,Kimel2005}.
For this reason, MLD can only distinguish between antiferromagnetic domains with perpendicular orientation. 
{In the case of near threshold excitation of photoelectrons,  related magnetic linear and circular dichroism effects have been observed~\cite{Marx2000,Nakagawa2006,Hild2009,Nakagawa2012}.}
In photoemission spectroscopy a similar effect can be observed~\cite{Hillebrecht1995}. 
X-ray magnetic linear dichroism
photoemission electron microscopy (XMLD-PEEM) was successfully used to observe AFM domains
in a wide range of materials~\cite{Hillebrecht1995,Stohr1999,Nolting2000,Krug2008,Baldrati2019}, including both 
CuMnAs~\cite{Wadley2016,Wadley2018} and Mn$_2$Au thin films~\cite{Sapozhnik2018,Grigorev2021}. 

We perform MLD photoemission microscopy by acquiring two data sets, exciting with linearly polarized laser light (6.4~eV) parallel (p) and perpendicular (s) to the reflection plane of the laser beam.

The photoemission intensities for both measurements, integrated over the field of view, have first been normalized to each other.
Then we calculate the spatial distribution of the MLD 
asymmetry
$A_{\rm MLD}=[I_p(E_B,x,y)-I_s(E_B,x,y)]/[I_p(E_B,x,y)+I_s(E_B,x,y)]$.
Other contributions to the contrast, such as work function contrast, topographical contrast, impurities, and detector function are largely eliminated. 
Figs.~\ref{Fig1}(b,c) show the resulting magnetic contrast obtained at a binding energy ($E_B$) of $0.6$~eV. 
The color code for the corresponding N{\'e}el vector alignment
is indicated as an inset in Fig.~\ref{Fig1}(d).

\begin{figure*}{h}
\includegraphics[width=0.9 \textwidth]{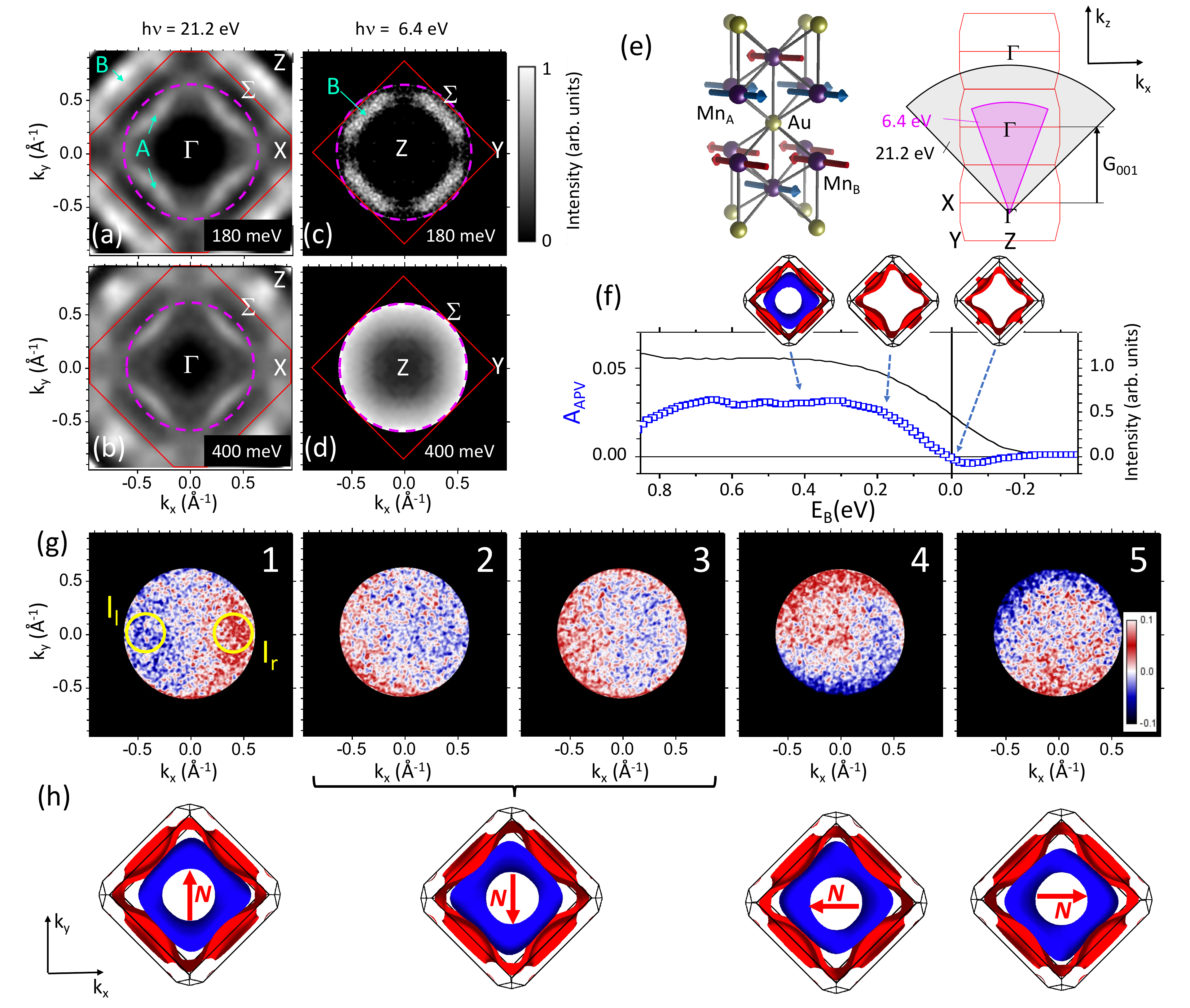}
\caption{\label{Fig2} 
(a,b) Momentum imaging with 21.2~eV photon energy, averaged over many domains.
Constant energy section at a binding energy of (a) $E_B=0.18$~eV and (b) $E_B=0.4$~eV of the photoemission intensity $I(E_B,k_x,k_y)$, corresponding to a 
cut through the constant energy surface near the $\Gamma$-X-$\Sigma$ plane as indicated in (e).
(c,d) Similar data for excitation with 6.4~eV. In this case, the momentum images correspond
to cuts through the constant energy surfaces near the Z-Y-$\Sigma$ plane (c,e).
Red lines indicate the corresponding Brillouin zone boundaries. 
The low photon energy (6.4~eV) restricts the field-of-view in momentum space to the violet dashed-circle (photoemission horizon)
that is shown also in 
(a,b) for comparison. 
(e) Schematic representation of crystal lattice and repeated Brillouin zone scheme in the $k_z$-$k_x$ plane,
indicating the $k_z$ values and observable 
$k_\parallel$ range for 21.2~eV and 6.4~eV excitation (spherical surface section appearing circular in top view).
(f) Photoemission intensity (black line) and $A_{\rm APV}$ (squares, see text for definition) 
for 6.4~eV excitation versus binding energy measured at the spatial region 1 in Fig.~\ref{Fig1}(c). 
Also shown above the graph are the calculated bandstructures at binding energy $0.0$, $0.2$, and $0.4$ eV, as in Fig.~\ref{FigConcept}(b).
(g) Antiferromagnetic parity violation measured at a binding energy of 0.4~eV
for the five regions 1-5 defined in Fig.~\ref{Fig1}(c). The corresponding asymmetry scale is indicated by the color scale bar in the inset of (g-5).
The yellow circles in (g-1) define the momentum areas used to determine the asymmetry values versus binding energy shown in (f).
(h) Calculated constant energy surfaces in $k$-space at $E_B=0.4$~eV for the indicated N{\'e}el vectors ${\bf N}$ (red arrows).
}
\vspace{-0,5 cm}
\end{figure*}

The result shows 
that it is indeed possible in this mode 
to distinguish domains with 
$\bf{N}$ parallel or perpendicular to $x$, but not whether
$\bf{N}$ is pointing up or down ~\cite{Sapozhnik2018}.

We note that $A_{\rm MLD}$ 
depends on $E_B$. 
Fig.~\ref{Fig1}(d)  depicts $A_{\rm MLD}(E_B)$
extracted from a single domain area in Fig.~\ref{Fig1}(c), revealing
a pronounced asymmetry peak near 
$E_B=0.6$~eV.
We attribute the peak to a 
resonant excitation of a spin-orbit split 
state 
at this $E_B$~\cite{Elmers2020}. 

We next examine the results of photoemission in momentum space, depicted in
Fig.~\ref{Fig2}. Assuming direct transitions into quasi-free-electron-like final states~\cite{Hufner2003,Reinert2005},
the 
final state momentum magnitude is given by: 
\begin{equation}\label{eq1}
	k_{\rm final} = (1/\hbar)\sqrt{2m_{eff}E_{\rm final}}; \;\; E_{\rm final} = h\nu-E_B+V_0^*,
\end{equation}
where $m_{eff}$ denotes the effective electron mass, $h\nu$ the photon energy, 
and $V^*_0\approx 10$~eV the inner potential with respect to the Fermi energy. 

Figs.~\ref{Fig2}(a,b) show results obtained for excitation with 21.2~eV photons using a hemispherical analyzer (see results for $E_B=0$ in Ref.~\onlinecite{Elmers2020} for comparison). 
In this case, the maximum value of the perpendicular momentum component is $k_z=2.86$~\AA$^{-1} = 1.9 \,G_{001}$.
The constant energy sections at low binding energies correspond to sections of the repeated Brillouin zone scheme
close to the  $\Gamma$-X-$\Sigma$ plane. Fig.~\ref{Fig2}(e) illustrates the actual spherical section.
This measurement averages over many domains with the N{\'e}el vector pointing along 
all four possible $\langle 110 \rangle$ directions. 
Correspondingly, the constant energy section is expected to display a four-fold symmetry.
Sections shown in Figs.~\ref{Fig2}(a-d) are symmetrized, accordingly.

Figs.~\ref{Fig2}(c,d) show results obtained for excitation with 6.4~eV photons using ToFMM for comparison. 
For this measurement, the photoemission intensity stems from a region of interest of 4.5~$\mu$m diameter on the sample surface,
which still averages over several domains [see Fig.~\ref{Fig1}(b,c)].
Here, the lower photon energy leads to a maximum perpendicular momentum at $k_x=k_y=0$ of
$k_z=2.07$~\AA$^{-1} = 1.4\,G_{001}$.
Thus, the probed section in the repeated Brillouin zone scheme is now close to the Z-$\Sigma$-Y plane, as sketched in Fig.~\ref{Fig2}(e).
The lower photon energy restricts the observable parallel momentum to
$k_{\parallel}<0.6$~\AA$^{-1}$, indicating the photoemission horizon.
Using the time-of-flight detection, we simultaneously acquire a data array
$I(E_B,k_x,k_y)$ where constant energy sections can, during post-processing, be selected for different binding energies.

For $E_B=0.18$~eV, we find the features B in Fig.~\ref{Fig2}(b),
which originate from the same band as observed for 21.2~eV in Fig.~\ref{Fig2}(a) closer to the Z point.
At the higher binding energy  $E_B=0.4$~eV this band seemingly broadens [Fig.~\ref{Fig2}(b)]
and leads to the circular-shaped high intensity for 6.4~eV excitation in Fig.~\ref{Fig2}(d) near the photoemission horizon.
These results confirm that $I(E_B,k_x,k_y)$ probes the spectral function (modulated by photoemission transition probabilities)
also at this low photon energy, 
despite the fact that the final state is less free-electron like, which 
leads to an integration over a finite $k_z$-interval.

Next, we inserted a 10~$\mu$m field aperture, 
mounted on a piezo-adjustable holder, at the position of a Gaussian image.
The latter is magnified by a factor of 11 at this position. 
Thus, the field aperture restricts electron paths to a circular shaped region of interest with a diameter of 0.9~$\mu$m.
The electron optics downstream from the field aperture is then switched to momentum mode such that
only electrons stemming from the selected region of interest contribute to the momentum image.

Similar data arrays $I_i(E_B,k_x,k_y)$ were measured for the five positions (1-5) of the field aperture indicated in Fig.~\ref{Fig1}(c).
These positions have been chosen to cover the two (red and blue) antiferromagnetic domains shown in Fig.~\ref{Fig1}(c).
To reveal the asymmetry in the intensity distribution in momentum space arising from the domains in the aperture in the spatial position $i$
we rescale the intensity signal according to
\begin{equation}
\tilde{I}_{i}(E_B,k_x,k_y)=\frac{I_i(E_B,k_x,k_y)-\overline{I(E_B,k_x,k_y)}}{\overline{I(E_B,k_x,k_y)}},
\end{equation}
where the mean intensity distribution in momentum space is given by 
$\overline{I(E_B,k_x,k_y)}=\sum_i{I_i(E_B,k_x,k_y)}/N$.
In the absence of magnetic order $\tilde{I}_{i}$ will be zero.
Fig.~\ref{Fig2}(g-1) depicts this asymmetry distribution for position 1 at a binding energy $E_B=0.4$~eV,
revealing a left/right asymmetry.
To determine its dependence on $E_B$
the asymmetry is averaged over the indicated left and right circular areas,
$\tilde{I}_l(E_B)$ and $\tilde{I}_r(E_B)$. 
We then define the APV as 
${\rm A_{\rm APV}} \equiv  \tilde{I}_l(E_B) - \tilde{I}_r(E_B)$,
depicted in Fig.~\ref{Fig2}(f) by open blue squares. 
We find a broad maximum value of $A_{\rm{APV}}=0.03$ between $E_B=0.2$~eV and 0.6~eV. 
This is consistent with the DFT bandstructure calculations shown
above panel (f) at energies $0.0, 0.2$, and $0.4$ eV. At $0.0$ eV energy there is no anisotropy.

At position 1 the magnetic linear dichroism observed in PEEM mode [Fig.~\ref{Fig1}(c)]
indicates a N{\'e}el vector alignment parallel to the $y$-axis.
According to the theoretical prediction this should lead to an APV ($E(k)\neq E(-k)$)
perpendicular to the N{\'e}el vector, i.e. along the $x$-axis,
in agreement with the experimental observation.
At positions 2 and 3 the APV is also oriented along the $x$-axis but reversed [Fig.~\ref{Fig2}(g-2,g-3)]. 
The reversed APV thus indicates antiparallel N{\'e}el vectors in regions 1 as compared to 2 and 3.
Positions 4 and 5 show an asymmetry along the $y$-axis [Fig.~\ref{Fig2}(g-4,g-5)], revealing a
N{\'e}el vector orientation parallel to the $x$-axis.
This is in agreement with the magnetic linear dichroism observed in the spatial domain distribution, too,
where these regions appear blue in Fig.~\ref{Fig1}(c).
The reversed asymmetry between region 4 and region 5 indicates an antiparallel orientation
of the N{\'e}el vector in these two domains.
The asymmetry distributions observed for the five regions of interest defined by the position of the field aperture
thus reveal all four possible orientations of the N{\'e}el vector, 
an information impossible to obtain with conventional ARPES (without sub-$\mu$m spatial resolution).

The calculated constant energy surfaces (see Methods) shown in Figs.~\ref{Fig2}(h)
reveal the origin of the experimentally observed APV.
The APV results in a pronounced extension
of the inner toroidal (blue) surface in the direction perpendicular to the N{\'e}el vector.
More specifically, the extension occurs to the right with respect to the N{\'e}el vector orientation.
Figs.~\ref{Fig2}(h) depict the  N{\'e}el vector orientations in accordance with the observed asymmetry in each of the measured regions.
The toroidal  surface, completely lying inside the first Brillouin zone, is barely 
touched by the nominal section probed with 6.4~eV excitation [see Fig.~\ref{Fig2}(e)],
explaining the absence of clear direct transition features in the intensity maps.
On the other hand, low photon energies lead to a probed spherical section integrated 
over a more extended $k_z$ interval.
The extension of the toroidal surfaces in Figs.~\ref{Fig2}(h) then
result in increased photoemission intensities stemming from averaged
direct 
transitions.
Thus, theory provides a direct link between the experimentally observed asymmetry
and the broken parity symmetry of energies. 

The connection of the DFT calculations and the sub-$\mu$-ToFMM technique gives us then a novel direct imaging technique 
that resolves the {\it direction} of the N\'eel vector domains, rather than only their alignment.
Although similar information could in principle be obtained by the second-harmonic generation \cite{Fiebig1994}, 
this N\'eel vector imaging technique has proven very challenging in metallic systems, being most successful in insulating magnetoelectric materials. 
The method is viable in principle in many antiferromagnets described by one of the 21 $\cal{PT}$ symmetric magnetic point groups, which 
account for a large fraction (17\%) of all magnetic point groups\cite{Engel2021}. 

In summary, we have directly observed  an APV, $E(k) \neq E(-k)$, in the collinear antiferromagnet Mn$_2$Au. 
The APV is caused by the combined effect of the  collinear magnetic structure of Mn$_2$Au,
where the two magnetic sublattices are connected {\it via} inversion symmetry, lifted by the magnetic moments,  and the spin-orbit coupling.
In addition, we have demonstrated that this APV in combination with the sub-$\mu$-ToFMM technique, 
allows the identification of 
directional N\'eel vector domains.

This work was funded by the Deutsche Forschungsgemeinschaft (DFG, German Research Foundation) Grant No. TRR 173 268565370 (projects A01, A02, A03, and A05), by the BMBF 
(projects 05K16UM1 and 05K19UM2), by the EU FET Open RIA Grant no. 766566, and by the Grant Agency of the Czech Republic grant no. 19-28375X.
Sincere thanks are due to A. Oelsner (Surface Concept GmbH) for continuous support.


\begin{thebibliography}{36}%
\makeatletter
\providecommand \@ifxundefined [1]{%
 \@ifx{#1\undefined}
}%
\providecommand \@ifnum [1]{%
 \ifnum #1\expandafter \@firstoftwo
 \else \expandafter \@secondoftwo
 \fi
}%
\providecommand \@ifx [1]{%
 \ifx #1\expandafter \@firstoftwo
 \else \expandafter \@secondoftwo
 \fi
}%
\providecommand \natexlab [1]{#1}%
\providecommand \enquote  [1]{``#1''}%
\providecommand \bibnamefont  [1]{#1}%
\providecommand \bibfnamefont [1]{#1}%
\providecommand \citenamefont [1]{#1}%
\providecommand \href@noop [0]{\@secondoftwo}%
\providecommand \href [0]{\begingroup \@sanitize@url \@href}%
\providecommand \@href[1]{\@@startlink{#1}\@@href}%
\providecommand \@@href[1]{\endgroup#1\@@endlink}%
\providecommand \@sanitize@url [0]{\catcode `\\12\catcode `\$12\catcode
  `\&12\catcode `\#12\catcode `\^12\catcode `\_12\catcode `\%12\relax}%
\providecommand \@@startlink[1]{}%
\providecommand \@@endlink[0]{}%
\providecommand \url  [0]{\begingroup\@sanitize@url \@url }%
\providecommand \@url [1]{\endgroup\@href {#1}{\urlprefix }}%
\providecommand \urlprefix  [0]{URL }%
\providecommand \Eprint [0]{\href }%
\providecommand \doibase [0]{http://dx.doi.org/}%
\providecommand \selectlanguage [0]{\@gobble}%
\providecommand \bibinfo  [0]{\@secondoftwo}%
\providecommand \bibfield  [0]{\@secondoftwo}%
\providecommand \translation [1]{[#1]}%
\providecommand \BibitemOpen [0]{}%
\providecommand \bibitemStop [0]{}%
\providecommand \bibitemNoStop [0]{.\EOS\space}%
\providecommand \EOS [0]{\spacefactor3000\relax}%
\providecommand \BibitemShut  [1]{\csname bibitem#1\endcsname}%
\let\auto@bib@innerbib\@empty
\bibitem [{\citenamefont {Sobota}\ \emph {et~al.}(2021)\citenamefont {Sobota},
  \citenamefont {He},\ and\ \citenamefont {Shen}}]{Sobota2021}%
  \BibitemOpen
  \bibfield  {author} {\bibinfo {author} {\bibfnamefont {J.~A.}\ \bibnamefont
  {Sobota}}, \bibinfo {author} {\bibfnamefont {Y.}~\bibnamefont {He}}, \ and\
  \bibinfo {author} {\bibfnamefont {Z.-X.}\ \bibnamefont {Shen}},\ }\href
  {\doibase 10.1103/RevModPhys.93.025006} {\bibfield  {journal} {\bibinfo
  {journal} {Rev. Mod. Phys.}\ }\textbf {\bibinfo {volume} {93}},\ \bibinfo
  {pages} {025006} (\bibinfo {year} {2021})}\BibitemShut {NoStop}%
\bibitem [{\citenamefont {{\v{S}}mejkal}\ \emph {et~al.}(2021)\citenamefont
  {{\v{S}}mejkal}, \citenamefont {Sinova},\ and\ \citenamefont
  {Jungwirth}}]{Smejkal2021a}%
  \BibitemOpen
  \bibfield  {author} {\bibinfo {author} {\bibfnamefont {L.}~\bibnamefont
  {{\v{S}}mejkal}}, \bibinfo {author} {\bibfnamefont {J.}~\bibnamefont
  {Sinova}}, \ and\ \bibinfo {author} {\bibfnamefont {T.}~\bibnamefont
  {Jungwirth}},\ }\href {http://arxiv.org/abs/2105.05820} {\  (\bibinfo {year}
  {2021})},\ \Eprint {http://arxiv.org/abs/2105.05820} {arXiv:2105.05820}
  \BibitemShut {NoStop}%
\bibitem [{\citenamefont {Smejkal}\ \emph {et~al.}(2017)\citenamefont
  {Smejkal}, \citenamefont {Zelezny}, \citenamefont {Sinova},\ and\
  \citenamefont {Jungwirth}}]{Smejkal2017}%
  \BibitemOpen
  \bibfield  {author} {\bibinfo {author} {\bibfnamefont {L.}~\bibnamefont
  {Smejkal}}, \bibinfo {author} {\bibfnamefont {J.}~\bibnamefont {Zelezny}},
  \bibinfo {author} {\bibfnamefont {J.}~\bibnamefont {Sinova}}, \ and\ \bibinfo
  {author} {\bibfnamefont {T.}~\bibnamefont {Jungwirth}},\ }\href {\doibase
  {10.1103/PhysRevLett.118.106402}} {\bibfield  {journal} {\bibinfo  {journal}
  {Phys. Rev. Lett.}\ }\textbf {\bibinfo {volume} {{118}}},\ \bibinfo {pages}
  {{106402}} (\bibinfo {year} {{2017}})}\BibitemShut {NoStop}%
\bibitem [{\citenamefont {Hayami}\ \emph {et~al.}(2018)\citenamefont {Hayami},
  \citenamefont {Yatsushiro}, \citenamefont {Yanagi},\ and\ \citenamefont
  {Kusunose}}]{Hayami2018}%
  \BibitemOpen
  \bibfield  {author} {\bibinfo {author} {\bibfnamefont {S.}~\bibnamefont
  {Hayami}}, \bibinfo {author} {\bibfnamefont {M.}~\bibnamefont {Yatsushiro}},
  \bibinfo {author} {\bibfnamefont {Y.}~\bibnamefont {Yanagi}}, \ and\ \bibinfo
  {author} {\bibfnamefont {H.}~\bibnamefont {Kusunose}},\ }\href {\doibase
  10.1103/PhysRevB.98.165110} {\bibfield  {journal} {\bibinfo  {journal} {Phys.
  Rev. B}\ }\textbf {\bibinfo {volume} {98}},\ \bibinfo {pages} {165110}
  (\bibinfo {year} {2018})}\BibitemShut {NoStop}%
\bibitem [{\citenamefont {Watanabe}\ and\ \citenamefont
  {Yanase}(2021)}]{Watanabe2020}%
  \BibitemOpen
  \bibfield  {author} {\bibinfo {author} {\bibfnamefont {H.}~\bibnamefont
  {Watanabe}}\ and\ \bibinfo {author} {\bibfnamefont {Y.}~\bibnamefont
  {Yanase}},\ }\href {\doibase 10.1103/PhysRevX.11.011001} {\bibfield
  {journal} {\bibinfo  {journal} {Physical Review X}\ }\textbf {\bibinfo
  {volume} {11}},\ \bibinfo {pages} {011001} (\bibinfo {year}
  {2021})}\BibitemShut {NoStop}%
\bibitem [{\citenamefont {Elmers}\ \emph {et~al.}(2020)\citenamefont {Elmers},
  \citenamefont {Chernov}, \citenamefont {DSouza}, \citenamefont
  {Bommanaboyena}, \citenamefont {Bodnar}, \citenamefont {Medjanik},
  \citenamefont {Babenkov}, \citenamefont {Fedchenko}, \citenamefont
  {Vasilyev}, \citenamefont {Agustsson}, \citenamefont {Schlueter},
  \citenamefont {Gloskovskii}, \citenamefont {Matveyev}, \citenamefont
  {Strocov}, \citenamefont {Skourski}, \citenamefont {{\v{S}}mejkal},
  \citenamefont {Sinova}, \citenamefont {Minar}, \citenamefont {Klaeui},
  \citenamefont {Schoenhense},\ and\ \citenamefont {Jourdan}}]{Elmers2020}%
  \BibitemOpen
  \bibfield  {author} {\bibinfo {author} {\bibfnamefont {H.~J.}\ \bibnamefont
  {Elmers}}, \bibinfo {author} {\bibfnamefont {S.~V.}\ \bibnamefont {Chernov}},
  \bibinfo {author} {\bibfnamefont {S.~W.}\ \bibnamefont {DSouza}}, \bibinfo
  {author} {\bibfnamefont {S.~P.}\ \bibnamefont {Bommanaboyena}}, \bibinfo
  {author} {\bibfnamefont {S.~Y.}\ \bibnamefont {Bodnar}}, \bibinfo {author}
  {\bibfnamefont {K.}~\bibnamefont {Medjanik}}, \bibinfo {author}
  {\bibfnamefont {S.}~\bibnamefont {Babenkov}}, \bibinfo {author}
  {\bibfnamefont {O.}~\bibnamefont {Fedchenko}}, \bibinfo {author}
  {\bibfnamefont {D.}~\bibnamefont {Vasilyev}}, \bibinfo {author}
  {\bibfnamefont {S.~Y.}\ \bibnamefont {Agustsson}}, \bibinfo {author}
  {\bibfnamefont {C.}~\bibnamefont {Schlueter}}, \bibinfo {author}
  {\bibfnamefont {A.}~\bibnamefont {Gloskovskii}}, \bibinfo {author}
  {\bibfnamefont {Y.}~\bibnamefont {Matveyev}}, \bibinfo {author}
  {\bibfnamefont {V.~N.}\ \bibnamefont {Strocov}}, \bibinfo {author}
  {\bibfnamefont {Y.}~\bibnamefont {Skourski}}, \bibinfo {author}
  {\bibfnamefont {L.}~\bibnamefont {{\v{S}}mejkal}}, \bibinfo {author}
  {\bibfnamefont {J.}~\bibnamefont {Sinova}}, \bibinfo {author} {\bibfnamefont
  {J.}~\bibnamefont {Minar}}, \bibinfo {author} {\bibfnamefont
  {M.}~\bibnamefont {Klaeui}}, \bibinfo {author} {\bibfnamefont
  {G.}~\bibnamefont {Schoenhense}}, \ and\ \bibinfo {author} {\bibfnamefont
  {M.}~\bibnamefont {Jourdan}},\ }\href {\doibase 10.1021/acsnano.0c08215}
  {\bibfield  {journal} {\bibinfo  {journal} {ACS Nano}\ }\textbf {\bibinfo
  {volume} {14}},\ \bibinfo {pages} {17554} (\bibinfo {year} {2020})},\
  \bibinfo {note} {pMID: 33236903},\ \Eprint
  {http://arxiv.org/abs/https://doi.org/10.1021/acsnano.0c08215}
  {https://doi.org/10.1021/acsnano.0c08215} \BibitemShut {NoStop}%
\bibitem [{\citenamefont {Hayami}\ \emph {et~al.}(2020)\citenamefont {Hayami},
  \citenamefont {Yanagi},\ and\ \citenamefont {Kusunose}}]{Hayami2020}%
  \BibitemOpen
  \bibfield  {author} {\bibinfo {author} {\bibfnamefont {S.}~\bibnamefont
  {Hayami}}, \bibinfo {author} {\bibfnamefont {Y.}~\bibnamefont {Yanagi}}, \
  and\ \bibinfo {author} {\bibfnamefont {H.}~\bibnamefont {Kusunose}},\ }\href
  {\doibase 10.1103/PhysRevB.102.144441} {\bibfield  {journal} {\bibinfo
  {journal} {Physical Review B}\ }\textbf {\bibinfo {volume} {102}},\ \bibinfo
  {pages} {144441} (\bibinfo {year} {2020})}\BibitemShut {NoStop}%
\bibitem [{\citenamefont {Carbone}\ \emph {et~al.}(2016)\citenamefont
  {Carbone}, \citenamefont {Moras}, \citenamefont {Sheverdyaeva}, \citenamefont
  {Pacil\'e}, \citenamefont {Papagno}, \citenamefont {Ferrari}, \citenamefont
  {Topwal}, \citenamefont {Vescovo}, \citenamefont {Bihlmayer}, \citenamefont
  {Freimuth}, \citenamefont {Mokrousov},\ and\ \citenamefont
  {Bl\"ugel}}]{Carbone2016}%
  \BibitemOpen
  \bibfield  {author} {\bibinfo {author} {\bibfnamefont {C.}~\bibnamefont
  {Carbone}}, \bibinfo {author} {\bibfnamefont {P.}~\bibnamefont {Moras}},
  \bibinfo {author} {\bibfnamefont {P.~M.}\ \bibnamefont {Sheverdyaeva}},
  \bibinfo {author} {\bibfnamefont {D.}~\bibnamefont {Pacil\'e}}, \bibinfo
  {author} {\bibfnamefont {M.}~\bibnamefont {Papagno}}, \bibinfo {author}
  {\bibfnamefont {L.}~\bibnamefont {Ferrari}}, \bibinfo {author} {\bibfnamefont
  {D.}~\bibnamefont {Topwal}}, \bibinfo {author} {\bibfnamefont
  {E.}~\bibnamefont {Vescovo}}, \bibinfo {author} {\bibfnamefont
  {G.}~\bibnamefont {Bihlmayer}}, \bibinfo {author} {\bibfnamefont
  {F.}~\bibnamefont {Freimuth}}, \bibinfo {author} {\bibfnamefont
  {Y.}~\bibnamefont {Mokrousov}}, \ and\ \bibinfo {author} {\bibfnamefont
  {S.}~\bibnamefont {Bl\"ugel}},\ }\href {\doibase 10.1103/PhysRevB.93.125409}
  {\bibfield  {journal} {\bibinfo  {journal} {Phys. Rev. B}\ }\textbf {\bibinfo
  {volume} {93}},\ \bibinfo {pages} {125409} (\bibinfo {year}
  {2016})}\BibitemShut {NoStop}%
\bibitem [{\citenamefont {Zelezny}\ \emph {et~al.}(2014)\citenamefont
  {Zelezny}, \citenamefont {Gao}, \citenamefont {Vyborny}, \citenamefont
  {Zemen}, \citenamefont {Masek}, \citenamefont {Manchon}, \citenamefont
  {Wunderlich}, \citenamefont {Sinova},\ and\ \citenamefont
  {Jungwirth}}]{Zelezny2014}%
  \BibitemOpen
  \bibfield  {author} {\bibinfo {author} {\bibfnamefont {J.}~\bibnamefont
  {Zelezny}}, \bibinfo {author} {\bibfnamefont {H.}~\bibnamefont {Gao}},
  \bibinfo {author} {\bibfnamefont {K.}~\bibnamefont {Vyborny}}, \bibinfo
  {author} {\bibfnamefont {J.}~\bibnamefont {Zemen}}, \bibinfo {author}
  {\bibfnamefont {J.}~\bibnamefont {Masek}}, \bibinfo {author} {\bibfnamefont
  {A.}~\bibnamefont {Manchon}}, \bibinfo {author} {\bibfnamefont
  {J.}~\bibnamefont {Wunderlich}}, \bibinfo {author} {\bibfnamefont
  {J.}~\bibnamefont {Sinova}}, \ and\ \bibinfo {author} {\bibfnamefont
  {T.}~\bibnamefont {Jungwirth}},\ }\href {\doibase
  {10.1103/PhysRevLett.113.157201}} {\bibfield  {journal} {\bibinfo  {journal}
  {Phys. Rev. Lett.}\ }\textbf {\bibinfo {volume} {{113}}},\ \bibinfo {pages}
  {{157201}} (\bibinfo {year} {{2014}})}\BibitemShut {NoStop}%
\bibitem [{\citenamefont {Wadley}\ \emph {et~al.}(2016)\citenamefont {Wadley},
  \citenamefont {Howells}, \citenamefont {Zelezny}, \citenamefont {Andrews},
  \citenamefont {Hills}, \citenamefont {Campion}, \citenamefont {Novak},
  \citenamefont {Olejnik}, \citenamefont {Maccherozzi}, \citenamefont {Dhesi},
  \citenamefont {Martin}, \citenamefont {Wagner}, \citenamefont {Wunderlich},
  \citenamefont {Freimuth}, \citenamefont {Mokrousov}, \citenamefont {Kunes},
  \citenamefont {Chauhan}, \citenamefont {Grzybowski}, \citenamefont
  {Rushforth},\ and\ \citenamefont {Edmonds}}]{Wadley2016}%
  \BibitemOpen
  \bibfield  {author} {\bibinfo {author} {\bibfnamefont {P.}~\bibnamefont
  {Wadley}}, \bibinfo {author} {\bibfnamefont {B.}~\bibnamefont {Howells}},
  \bibinfo {author} {\bibfnamefont {J.}~\bibnamefont {Zelezny}}, \bibinfo
  {author} {\bibfnamefont {C.}~\bibnamefont {Andrews}}, \bibinfo {author}
  {\bibfnamefont {V.}~\bibnamefont {Hills}}, \bibinfo {author} {\bibfnamefont
  {R.~P.}\ \bibnamefont {Campion}}, \bibinfo {author} {\bibfnamefont
  {V.}~\bibnamefont {Novak}}, \bibinfo {author} {\bibfnamefont
  {K.}~\bibnamefont {Olejnik}}, \bibinfo {author} {\bibfnamefont
  {F.}~\bibnamefont {Maccherozzi}}, \bibinfo {author} {\bibfnamefont {S.~S.}\
  \bibnamefont {Dhesi}}, \bibinfo {author} {\bibfnamefont {S.~Y.}\ \bibnamefont
  {Martin}}, \bibinfo {author} {\bibfnamefont {T.}~\bibnamefont {Wagner}},
  \bibinfo {author} {\bibfnamefont {J.}~\bibnamefont {Wunderlich}}, \bibinfo
  {author} {\bibfnamefont {F.}~\bibnamefont {Freimuth}}, \bibinfo {author}
  {\bibfnamefont {Y.}~\bibnamefont {Mokrousov}}, \bibinfo {author}
  {\bibfnamefont {J.}~\bibnamefont {Kunes}}, \bibinfo {author} {\bibfnamefont
  {J.~S.}\ \bibnamefont {Chauhan}}, \bibinfo {author} {\bibfnamefont {M.~J.}\
  \bibnamefont {Grzybowski}}, \bibinfo {author} {\bibfnamefont {A.~W.}\
  \bibnamefont {Rushforth}}, \ and\ \bibinfo {author} {\bibfnamefont
  {K.~W.~{\it et al}.}\ \bibnamefont {Edmonds}},\ }\href {\doibase
  {10.1126/science.aab1031}} {\bibfield  {journal} {\bibinfo  {journal}
  {{Science}}\ }\textbf {\bibinfo {volume} {{351}}},\ \bibinfo {pages} {{587}}
  (\bibinfo {year} {{2016}})}\BibitemShut {NoStop}%
\bibitem [{\citenamefont {Bodnar}\ \emph {et~al.}(2018)\citenamefont {Bodnar},
  \citenamefont {Smejkal}, \citenamefont {Turek}, \citenamefont {Jungwirth},
  \citenamefont {Gomonay}, \citenamefont {Sinova}, \citenamefont {Sapozhnik},
  \citenamefont {Elmers}, \citenamefont {Kl{\"a}ui},\ and\ \citenamefont
  {Jourdan}}]{Bodnar2018}%
  \BibitemOpen
  \bibfield  {author} {\bibinfo {author} {\bibfnamefont {S.~Y.}\ \bibnamefont
  {Bodnar}}, \bibinfo {author} {\bibfnamefont {L.}~\bibnamefont {Smejkal}},
  \bibinfo {author} {\bibfnamefont {I.}~\bibnamefont {Turek}}, \bibinfo
  {author} {\bibfnamefont {T.}~\bibnamefont {Jungwirth}}, \bibinfo {author}
  {\bibfnamefont {O.}~\bibnamefont {Gomonay}}, \bibinfo {author} {\bibfnamefont
  {J.}~\bibnamefont {Sinova}}, \bibinfo {author} {\bibfnamefont {A.~A.}\
  \bibnamefont {Sapozhnik}}, \bibinfo {author} {\bibfnamefont {H.~J.}\
  \bibnamefont {Elmers}}, \bibinfo {author} {\bibfnamefont {M.}~\bibnamefont
  {Kl{\"a}ui}}, \ and\ \bibinfo {author} {\bibfnamefont {M.}~\bibnamefont
  {Jourdan}},\ }\href@noop {} {\bibfield  {journal} {\bibinfo  {journal} {Nat.
  Comm.}\ }\textbf {\bibinfo {volume} {9}},\ \bibinfo {pages} {348} (\bibinfo
  {year} {2018})}\BibitemShut {NoStop}%
\bibitem [{\citenamefont {Godinho}\ \emph {et~al.}(2018)\citenamefont
  {Godinho}, \citenamefont {Reichlova}, \citenamefont {Kriegner}, \citenamefont
  {Novak}, \citenamefont {Olejnik}, \citenamefont {Kaspar}, \citenamefont
  {Soban}, \citenamefont {Wadley}, \citenamefont {Campion}, \citenamefont
  {Otxoa}, \citenamefont {Roy}, \citenamefont {Zelezny}, \citenamefont
  {Jungwirth},\ and\ \citenamefont {Wunderlich}}]{Godinho2018}%
  \BibitemOpen
  \bibfield  {author} {\bibinfo {author} {\bibfnamefont {J.}~\bibnamefont
  {Godinho}}, \bibinfo {author} {\bibfnamefont {H.}~\bibnamefont {Reichlova}},
  \bibinfo {author} {\bibfnamefont {D.}~\bibnamefont {Kriegner}}, \bibinfo
  {author} {\bibfnamefont {V.}~\bibnamefont {Novak}}, \bibinfo {author}
  {\bibfnamefont {K.}~\bibnamefont {Olejnik}}, \bibinfo {author} {\bibfnamefont
  {Z.}~\bibnamefont {Kaspar}}, \bibinfo {author} {\bibfnamefont
  {Z.}~\bibnamefont {Soban}}, \bibinfo {author} {\bibfnamefont
  {P.}~\bibnamefont {Wadley}}, \bibinfo {author} {\bibfnamefont {R.~P.}\
  \bibnamefont {Campion}}, \bibinfo {author} {\bibfnamefont {R.~M.}\
  \bibnamefont {Otxoa}}, \bibinfo {author} {\bibfnamefont {P.~E.}\ \bibnamefont
  {Roy}}, \bibinfo {author} {\bibfnamefont {J.}~\bibnamefont {Zelezny}},
  \bibinfo {author} {\bibfnamefont {T.}~\bibnamefont {Jungwirth}}, \ and\
  \bibinfo {author} {\bibfnamefont {J.}~\bibnamefont {Wunderlich}},\ }\href
  {\doibase 10.1038/s41467-018-07092-2} {\bibfield  {journal} {\bibinfo
  {journal} {Nat. Commun.}\ }\textbf {\bibinfo {volume} {9}},\ \bibinfo {pages}
  {4686} (\bibinfo {year} {2018})},\ \Eprint {http://arxiv.org/abs/1806.02795}
  {arXiv:1806.02795} \BibitemShut {NoStop}%
\bibitem [{\citenamefont {Bommanaboyena}\ \emph {et~al.}(2021)\citenamefont
  {Bommanaboyena}, \citenamefont {Backes}, \citenamefont {Veiga}, \citenamefont
  {Dhesi}, \citenamefont {Niu}, \citenamefont {Sarpi}, \citenamefont
  {Denneulin}, \citenamefont {Kovacs}, \citenamefont {Mashoff}, \citenamefont
  {Gomonay}, \citenamefont {Sinova}, \citenamefont {Everschor-Sitte},
  \citenamefont {Sch{\"o}nke}, \citenamefont {Reeve}, \citenamefont
  {Kl{\"a}ui}, \citenamefont {Elmers},\ and\ \citenamefont
  {Jourdan}}]{Bommanaboyena2021}%
  \BibitemOpen
  \bibfield  {author} {\bibinfo {author} {\bibfnamefont {S.}~\bibnamefont
  {Bommanaboyena}}, \bibinfo {author} {\bibfnamefont {B.}~\bibnamefont
  {Backes}}, \bibinfo {author} {\bibfnamefont {L.~S.~I.}\ \bibnamefont
  {Veiga}}, \bibinfo {author} {\bibfnamefont {S.~S.}\ \bibnamefont {Dhesi}},
  \bibinfo {author} {\bibfnamefont {Y.~R.}\ \bibnamefont {Niu}}, \bibinfo
  {author} {\bibfnamefont {B.}~\bibnamefont {Sarpi}}, \bibinfo {author}
  {\bibfnamefont {T.}~\bibnamefont {Denneulin}}, \bibinfo {author}
  {\bibfnamefont {A.}~\bibnamefont {Kovacs}}, \bibinfo {author} {\bibfnamefont
  {T.}~\bibnamefont {Mashoff}}, \bibinfo {author} {\bibfnamefont
  {O.}~\bibnamefont {Gomonay}}, \bibinfo {author} {\bibfnamefont
  {J.}~\bibnamefont {Sinova}}, \bibinfo {author} {\bibfnamefont
  {K.}~\bibnamefont {Everschor-Sitte}}, \bibinfo {author} {\bibfnamefont
  {D.}~\bibnamefont {Sch{\"o}nke}}, \bibinfo {author} {\bibfnamefont {R.~M.}\
  \bibnamefont {Reeve}}, \bibinfo {author} {\bibfnamefont {M.}~\bibnamefont
  {Kl{\"a}ui}}, \bibinfo {author} {\bibfnamefont {H.-J.}\ \bibnamefont
  {Elmers}}, \ and\ \bibinfo {author} {\bibfnamefont {M.}~\bibnamefont
  {Jourdan}},\ }\href {arXiv:2106.02333v1} {\bibfield  {journal} {\bibinfo
  {journal} {Arxiv}\ } (\bibinfo {year} {2021})}\BibitemShut {NoStop}%
\bibitem [{\citenamefont {Jourdan}\ \emph {et~al.}(2015)\citenamefont
  {Jourdan}, \citenamefont {Braeuning}, \citenamefont {Sapozhnik},
  \citenamefont {Elmers}, \citenamefont {Zabel},\ and\ \citenamefont
  {Klaeui}}]{Jourdan2015}%
  \BibitemOpen
  \bibfield  {author} {\bibinfo {author} {\bibfnamefont {M.}~\bibnamefont
  {Jourdan}}, \bibinfo {author} {\bibfnamefont {H.}~\bibnamefont {Braeuning}},
  \bibinfo {author} {\bibfnamefont {A.}~\bibnamefont {Sapozhnik}}, \bibinfo
  {author} {\bibfnamefont {H.~J.}\ \bibnamefont {Elmers}}, \bibinfo {author}
  {\bibfnamefont {H.}~\bibnamefont {Zabel}}, \ and\ \bibinfo {author}
  {\bibfnamefont {M.}~\bibnamefont {Klaeui}},\ }\href {\doibase
  {10.1088/0022-3727/48/38/385001}} {\bibfield  {journal} {\bibinfo  {journal}
  {J. Phys. D: Appl. Phys.}\ }\textbf {\bibinfo {volume} {{48}}},\ \bibinfo
  {pages} {385001} (\bibinfo {year} {{2015}})}\BibitemShut {NoStop}%
\bibitem [{\citenamefont {Bohammaboyena}\ \emph {et~al.}(2020)\citenamefont
  {Bohammaboyena}, \citenamefont {Bergfeldt}, \citenamefont {Heller},
  \citenamefont {Kl{\"a}ui},\ and\ \citenamefont {Jourdan}}]{Satya2020}%
  \BibitemOpen
  \bibfield  {author} {\bibinfo {author} {\bibfnamefont {S.~P.}\ \bibnamefont
  {Bohammaboyena}}, \bibinfo {author} {\bibfnamefont {T.}~\bibnamefont
  {Bergfeldt}}, \bibinfo {author} {\bibfnamefont {R.}~\bibnamefont {Heller}},
  \bibinfo {author} {\bibfnamefont {M.}~\bibnamefont {Kl{\"a}ui}}, \ and\
  \bibinfo {author} {\bibfnamefont {M.}~\bibnamefont {Jourdan}},\ }\href
  {\doibase https://doi.org/10.1063/5.0009566} {\bibfield  {journal} {\bibinfo
  {journal} {{J. Appl. Phys.}}\ }\textbf {\bibinfo {volume} {127}},\ \bibinfo
  {pages} {243901} (\bibinfo {year} {2020})}\BibitemShut {NoStop}%
\bibitem [{\citenamefont {Sapozhnik}\ \emph {et~al.}(2018)\citenamefont
  {Sapozhnik}, \citenamefont {Filianina}, \citenamefont {Bodnar}, \citenamefont
  {Lamirand}, \citenamefont {Mawass}, \citenamefont {Skourski}, \citenamefont
  {Elmers}, \citenamefont {Zabel}, \citenamefont {Klaeui},\ and\ \citenamefont
  {Jourdan}}]{Sapozhnik2018}%
  \BibitemOpen
  \bibfield  {author} {\bibinfo {author} {\bibfnamefont {A.~A.}\ \bibnamefont
  {Sapozhnik}}, \bibinfo {author} {\bibfnamefont {M.}~\bibnamefont
  {Filianina}}, \bibinfo {author} {\bibfnamefont {S.~Y.}\ \bibnamefont
  {Bodnar}}, \bibinfo {author} {\bibfnamefont {A.}~\bibnamefont {Lamirand}},
  \bibinfo {author} {\bibfnamefont {M.-A.}\ \bibnamefont {Mawass}}, \bibinfo
  {author} {\bibfnamefont {Y.}~\bibnamefont {Skourski}}, \bibinfo {author}
  {\bibfnamefont {H.-J.}\ \bibnamefont {Elmers}}, \bibinfo {author}
  {\bibfnamefont {H.}~\bibnamefont {Zabel}}, \bibinfo {author} {\bibfnamefont
  {M.}~\bibnamefont {Klaeui}}, \ and\ \bibinfo {author} {\bibfnamefont
  {M.}~\bibnamefont {Jourdan}},\ }\href {\doibase {10.1103/PhysRevB.97.134429}}
  {\bibfield  {journal} {\bibinfo  {journal} {{Phys. Rev. B}}\ }\textbf
  {\bibinfo {volume} {{97}}},\ \bibinfo {pages} {{134429}} (\bibinfo {year}
  {{2018}})}\BibitemShut {NoStop}%
\bibitem [{\citenamefont {Hubert}\ and\ \citenamefont
  {Sch{\"a}fer}(1998)}]{Hubert1998}%
  \BibitemOpen
  \bibfield  {author} {\bibinfo {author} {\bibfnamefont {A.}~\bibnamefont
  {Hubert}}\ and\ \bibinfo {author} {\bibfnamefont {R.}~\bibnamefont
  {Sch{\"a}fer}},\ }\href {\doibase 10.1007/978-3-540-85054-0} {\emph {\bibinfo
  {title} {Magnetic Domains}}}\ (\bibinfo  {publisher} {Springer (Berlin)},\
  \bibinfo {year} {1998})\BibitemShut {NoStop}%
\bibitem [{\citenamefont {Kimel}\ \emph {et~al.}(2005)\citenamefont {Kimel},
  \citenamefont {Astakhov}, \citenamefont {Kirilyuk}, \citenamefont {Schott},
  \citenamefont {Karczewski}, \citenamefont {Ossau}, \citenamefont {Schmidt},
  \citenamefont {Molenkamp},\ and\ \citenamefont {Rasing}}]{Kimel2005}%
  \BibitemOpen
  \bibfield  {author} {\bibinfo {author} {\bibfnamefont {A.~V.}\ \bibnamefont
  {Kimel}}, \bibinfo {author} {\bibfnamefont {G.~V.}\ \bibnamefont {Astakhov}},
  \bibinfo {author} {\bibfnamefont {A.}~\bibnamefont {Kirilyuk}}, \bibinfo
  {author} {\bibfnamefont {G.~M.}\ \bibnamefont {Schott}}, \bibinfo {author}
  {\bibfnamefont {G.}~\bibnamefont {Karczewski}}, \bibinfo {author}
  {\bibfnamefont {W.}~\bibnamefont {Ossau}}, \bibinfo {author} {\bibfnamefont
  {G.}~\bibnamefont {Schmidt}}, \bibinfo {author} {\bibfnamefont {L.~W.}\
  \bibnamefont {Molenkamp}}, \ and\ \bibinfo {author} {\bibfnamefont
  {T.}~\bibnamefont {Rasing}},\ }\href {\doibase 10.1103/PhysRevLett.94.227203}
  {\bibfield  {journal} {\bibinfo  {journal} {Phys. Rev. Lett.}\ }\textbf
  {\bibinfo {volume} {94}},\ \bibinfo {pages} {227203} (\bibinfo {year}
  {2005})}\BibitemShut {NoStop}%
\bibitem [{\citenamefont {Marx}\ \emph {et~al.}(2000)\citenamefont {Marx},
  \citenamefont {Elmers},\ and\ \citenamefont {Sch\"onhense}}]{Marx2000}%
  \BibitemOpen
  \bibfield  {author} {\bibinfo {author} {\bibfnamefont {G.~K.~L.}\
  \bibnamefont {Marx}}, \bibinfo {author} {\bibfnamefont {H.~J.}\ \bibnamefont
  {Elmers}}, \ and\ \bibinfo {author} {\bibfnamefont {G.}~\bibnamefont
  {Sch\"onhense}},\ }\href {\doibase 10.1103/PhysRevLett.84.5888} {\bibfield
  {journal} {\bibinfo  {journal} {Phys. Rev. Lett.}\ }\textbf {\bibinfo
  {volume} {84}},\ \bibinfo {pages} {5888} (\bibinfo {year}
  {2000})}\BibitemShut {NoStop}%
\bibitem [{\citenamefont {Nakagawa}\ and\ \citenamefont
  {Yokoyama}(2006)}]{Nakagawa2006}%
  \BibitemOpen
  \bibfield  {author} {\bibinfo {author} {\bibfnamefont {T.}~\bibnamefont
  {Nakagawa}}\ and\ \bibinfo {author} {\bibfnamefont {T.}~\bibnamefont
  {Yokoyama}},\ }\href {\doibase 10.1103/PhysRevLett.96.237402} {\bibfield
  {journal} {\bibinfo  {journal} {Phys. Rev. Lett.}\ }\textbf {\bibinfo
  {volume} {96}},\ \bibinfo {pages} {237402} (\bibinfo {year}
  {2006})}\BibitemShut {NoStop}%
\bibitem [{\citenamefont {Hild}\ \emph {et~al.}(2009)\citenamefont {Hild},
  \citenamefont {Maul}, \citenamefont {Sch\"onhense}, \citenamefont {Elmers},
  \citenamefont {Amft},\ and\ \citenamefont {Oppeneer}}]{Hild2009}%
  \BibitemOpen
  \bibfield  {author} {\bibinfo {author} {\bibfnamefont {K.}~\bibnamefont
  {Hild}}, \bibinfo {author} {\bibfnamefont {J.}~\bibnamefont {Maul}}, \bibinfo
  {author} {\bibfnamefont {G.}~\bibnamefont {Sch\"onhense}}, \bibinfo {author}
  {\bibfnamefont {H.~J.}\ \bibnamefont {Elmers}}, \bibinfo {author}
  {\bibfnamefont {M.}~\bibnamefont {Amft}}, \ and\ \bibinfo {author}
  {\bibfnamefont {P.~M.}\ \bibnamefont {Oppeneer}},\ }\href {\doibase
  10.1103/PhysRevLett.102.057207} {\bibfield  {journal} {\bibinfo  {journal}
  {Phys. Rev. Lett.}\ }\textbf {\bibinfo {volume} {102}},\ \bibinfo {pages}
  {057207} (\bibinfo {year} {2009})}\BibitemShut {NoStop}%
\bibitem [{\citenamefont {Nakagawa}\ and\ \citenamefont
  {Yokoyama}(2012)}]{Nakagawa2012}%
  \BibitemOpen
  \bibfield  {author} {\bibinfo {author} {\bibfnamefont {T.}~\bibnamefont
  {Nakagawa}}\ and\ \bibinfo {author} {\bibfnamefont {T.}~\bibnamefont
  {Yokoyama}},\ }\href {\doibase 10.1016/j.elspec.2012.02.009} {\bibfield
  {journal} {\bibinfo  {journal} {J. Electr. Spectr. Rel. Phenom.}\ }\textbf
  {\bibinfo {volume} {185}},\ \bibinfo {pages} {356} (\bibinfo {year}
  {2012})}\BibitemShut {NoStop}%
\bibitem [{\citenamefont {Hillebrecht}\ \emph {et~al.}(1995)\citenamefont
  {Hillebrecht}, \citenamefont {Kinoshita}, \citenamefont {Spanke},
  \citenamefont {Dresselhaus}, \citenamefont {Roth}, \citenamefont {Rose},\
  and\ \citenamefont {Kisker}}]{Hillebrecht1995}%
  \BibitemOpen
  \bibfield  {author} {\bibinfo {author} {\bibfnamefont {F.~U.}\ \bibnamefont
  {Hillebrecht}}, \bibinfo {author} {\bibfnamefont {T.}~\bibnamefont
  {Kinoshita}}, \bibinfo {author} {\bibfnamefont {D.}~\bibnamefont {Spanke}},
  \bibinfo {author} {\bibfnamefont {J.}~\bibnamefont {Dresselhaus}}, \bibinfo
  {author} {\bibfnamefont {C.}~\bibnamefont {Roth}}, \bibinfo {author}
  {\bibfnamefont {H.~B.}\ \bibnamefont {Rose}}, \ and\ \bibinfo {author}
  {\bibfnamefont {E.}~\bibnamefont {Kisker}},\ }\href {\doibase
  10.1103/PhysRevLett.75.2224} {\bibfield  {journal} {\bibinfo  {journal}
  {Phys. Rev. Lett.}\ }\textbf {\bibinfo {volume} {75}},\ \bibinfo {pages}
  {2224} (\bibinfo {year} {1995})}\BibitemShut {NoStop}%
\bibitem [{\citenamefont {Stohr}\ \emph {et~al.}(1999)\citenamefont {Stohr},
  \citenamefont {Scholl}, \citenamefont {Regan}, \citenamefont {Anders},
  \citenamefont {Luning}, \citenamefont {Scheinfein}, \citenamefont {Padmore},\
  and\ \citenamefont {White}}]{Stohr1999}%
  \BibitemOpen
  \bibfield  {author} {\bibinfo {author} {\bibfnamefont {J.}~\bibnamefont
  {Stohr}}, \bibinfo {author} {\bibfnamefont {A.}~\bibnamefont {Scholl}},
  \bibinfo {author} {\bibfnamefont {T.}~\bibnamefont {Regan}}, \bibinfo
  {author} {\bibfnamefont {S.}~\bibnamefont {Anders}}, \bibinfo {author}
  {\bibfnamefont {J.}~\bibnamefont {Luning}}, \bibinfo {author} {\bibfnamefont
  {M.}~\bibnamefont {Scheinfein}}, \bibinfo {author} {\bibfnamefont
  {H.}~\bibnamefont {Padmore}}, \ and\ \bibinfo {author} {\bibfnamefont
  {R.}~\bibnamefont {White}},\ }\href {\doibase {10.1103/PhysRevLett.83.1862}}
  {\bibfield  {journal} {\bibinfo  {journal} {Phys. Rev. Lett.}\ }\textbf
  {\bibinfo {volume} {{83}}},\ \bibinfo {pages} {{1862}} (\bibinfo {year}
  {{1999}})}\BibitemShut {NoStop}%
\bibitem [{\citenamefont {Nolting}\ \emph {et~al.}(2000)\citenamefont
  {Nolting}, \citenamefont {Scholl}, \citenamefont {Stohr}, \citenamefont
  {Seo}, \citenamefont {Fompeyrine}, \citenamefont {Siegwart}, \citenamefont
  {Locquet}, \citenamefont {Anders}, \citenamefont {Luning}, \citenamefont
  {Fullerton}, \citenamefont {Toney}, \citenamefont {Scheinfein},\ and\
  \citenamefont {Padmore}}]{Nolting2000}%
  \BibitemOpen
  \bibfield  {author} {\bibinfo {author} {\bibfnamefont {F.}~\bibnamefont
  {Nolting}}, \bibinfo {author} {\bibfnamefont {A.}~\bibnamefont {Scholl}},
  \bibinfo {author} {\bibfnamefont {J.}~\bibnamefont {Stohr}}, \bibinfo
  {author} {\bibfnamefont {J.}~\bibnamefont {Seo}}, \bibinfo {author}
  {\bibfnamefont {J.}~\bibnamefont {Fompeyrine}}, \bibinfo {author}
  {\bibfnamefont {H.}~\bibnamefont {Siegwart}}, \bibinfo {author}
  {\bibfnamefont {J.}~\bibnamefont {Locquet}}, \bibinfo {author} {\bibfnamefont
  {S.}~\bibnamefont {Anders}}, \bibinfo {author} {\bibfnamefont
  {J.}~\bibnamefont {Luning}}, \bibinfo {author} {\bibfnamefont
  {E.}~\bibnamefont {Fullerton}}, \bibinfo {author} {\bibfnamefont
  {M.}~\bibnamefont {Toney}}, \bibinfo {author} {\bibfnamefont
  {M.}~\bibnamefont {Scheinfein}}, \ and\ \bibinfo {author} {\bibfnamefont
  {H.}~\bibnamefont {Padmore}},\ }\href {\doibase {10.1038/35015515}}
  {\bibfield  {journal} {\bibinfo  {journal} {Nature}\ }\textbf {\bibinfo
  {volume} {{405}}},\ \bibinfo {pages} {{767}} (\bibinfo {year}
  {{2000}})}\BibitemShut {NoStop}%
\bibitem [{\citenamefont {Krug}\ \emph {et~al.}(2008)\citenamefont {Krug},
  \citenamefont {Hillebrecht}, \citenamefont {Haverkort}, \citenamefont
  {Tanaka}, \citenamefont {Tjeng}, \citenamefont {Gomonay}, \citenamefont
  {Fraile-Rodrifguez}, \citenamefont {Nolting}, \citenamefont {Cramm},\ and\
  \citenamefont {Schneider}}]{Krug2008}%
  \BibitemOpen
  \bibfield  {author} {\bibinfo {author} {\bibfnamefont {I.~P.}\ \bibnamefont
  {Krug}}, \bibinfo {author} {\bibfnamefont {F.~U.}\ \bibnamefont
  {Hillebrecht}}, \bibinfo {author} {\bibfnamefont {M.~W.}\ \bibnamefont
  {Haverkort}}, \bibinfo {author} {\bibfnamefont {A.}~\bibnamefont {Tanaka}},
  \bibinfo {author} {\bibfnamefont {L.~H.}\ \bibnamefont {Tjeng}}, \bibinfo
  {author} {\bibfnamefont {H.}~\bibnamefont {Gomonay}}, \bibinfo {author}
  {\bibfnamefont {A.}~\bibnamefont {Fraile-Rodrifguez}}, \bibinfo {author}
  {\bibfnamefont {F.}~\bibnamefont {Nolting}}, \bibinfo {author} {\bibfnamefont
  {S.}~\bibnamefont {Cramm}}, \ and\ \bibinfo {author} {\bibfnamefont {C.~M.}\
  \bibnamefont {Schneider}},\ }\href {\doibase {10.1103/PhysRevB.78.064427}}
  {\bibfield  {journal} {\bibinfo  {journal} {Phys. Rev. B}\ }\textbf {\bibinfo
  {volume} {{78}}},\ \bibinfo {pages} {064427} (\bibinfo {year}
  {{2008}})}\BibitemShut {NoStop}%
\bibitem [{\citenamefont {Baldrati}\ \emph {et~al.}(2019)\citenamefont
  {Baldrati}, \citenamefont {Gomonay}, \citenamefont {Ross}, \citenamefont
  {Filianina}, \citenamefont {Lebrun}, \citenamefont {Ramos}, \citenamefont
  {Leveille}, \citenamefont {Fuhrmann}, \citenamefont {Forrest}, \citenamefont
  {Maccherozzi}, \citenamefont {Valencia}, \citenamefont {Kronast},
  \citenamefont {Saitoh}, \citenamefont {Sinova},\ and\ \citenamefont
  {Klaeui}}]{Baldrati2019}%
  \BibitemOpen
  \bibfield  {author} {\bibinfo {author} {\bibfnamefont {L.}~\bibnamefont
  {Baldrati}}, \bibinfo {author} {\bibfnamefont {O.}~\bibnamefont {Gomonay}},
  \bibinfo {author} {\bibfnamefont {A.}~\bibnamefont {Ross}}, \bibinfo {author}
  {\bibfnamefont {M.}~\bibnamefont {Filianina}}, \bibinfo {author}
  {\bibfnamefont {R.}~\bibnamefont {Lebrun}}, \bibinfo {author} {\bibfnamefont
  {R.}~\bibnamefont {Ramos}}, \bibinfo {author} {\bibfnamefont
  {C.}~\bibnamefont {Leveille}}, \bibinfo {author} {\bibfnamefont
  {F.}~\bibnamefont {Fuhrmann}}, \bibinfo {author} {\bibfnamefont {T.~R.}\
  \bibnamefont {Forrest}}, \bibinfo {author} {\bibfnamefont {F.}~\bibnamefont
  {Maccherozzi}}, \bibinfo {author} {\bibfnamefont {S.}~\bibnamefont
  {Valencia}}, \bibinfo {author} {\bibfnamefont {F.}~\bibnamefont {Kronast}},
  \bibinfo {author} {\bibfnamefont {E.}~\bibnamefont {Saitoh}}, \bibinfo
  {author} {\bibfnamefont {J.}~\bibnamefont {Sinova}}, \ and\ \bibinfo {author}
  {\bibfnamefont {M.}~\bibnamefont {Klaeui}},\ }\href {\doibase
  {10.1103/PhysRevLett.123.177201}} {\bibfield  {journal} {\bibinfo  {journal}
  {Phys. Rev. Lett.}\ }\textbf {\bibinfo {volume} {{123}}},\ \bibinfo {pages}
  {177201} (\bibinfo {year} {{2019}})}\BibitemShut {NoStop}%
\bibitem [{\citenamefont {Wadley}\ \emph {et~al.}(2018)\citenamefont {Wadley},
  \citenamefont {Reimers}, \citenamefont {Grzybowski}, \citenamefont {Andrews},
  \citenamefont {Wang}, \citenamefont {Chauhan}, \citenamefont {Gallagher},
  \citenamefont {Campion}, \citenamefont {Edmonds}, \citenamefont {Dhesi},
  \citenamefont {Maccherozzi}, \citenamefont {Novak}, \citenamefont
  {Wunderlich},\ and\ \citenamefont {Jungwirth}}]{Wadley2018}%
  \BibitemOpen
  \bibfield  {author} {\bibinfo {author} {\bibfnamefont {P.}~\bibnamefont
  {Wadley}}, \bibinfo {author} {\bibfnamefont {S.}~\bibnamefont {Reimers}},
  \bibinfo {author} {\bibfnamefont {M.~J.}\ \bibnamefont {Grzybowski}},
  \bibinfo {author} {\bibfnamefont {C.}~\bibnamefont {Andrews}}, \bibinfo
  {author} {\bibfnamefont {M.}~\bibnamefont {Wang}}, \bibinfo {author}
  {\bibfnamefont {J.~S.}\ \bibnamefont {Chauhan}}, \bibinfo {author}
  {\bibfnamefont {B.~L.}\ \bibnamefont {Gallagher}}, \bibinfo {author}
  {\bibfnamefont {R.~P.}\ \bibnamefont {Campion}}, \bibinfo {author}
  {\bibfnamefont {K.~W.}\ \bibnamefont {Edmonds}}, \bibinfo {author}
  {\bibfnamefont {S.~S.}\ \bibnamefont {Dhesi}}, \bibinfo {author}
  {\bibfnamefont {F.}~\bibnamefont {Maccherozzi}}, \bibinfo {author}
  {\bibfnamefont {V.}~\bibnamefont {Novak}}, \bibinfo {author} {\bibfnamefont
  {J.}~\bibnamefont {Wunderlich}}, \ and\ \bibinfo {author} {\bibfnamefont
  {T.}~\bibnamefont {Jungwirth}},\ }\href@noop {} {\bibfield  {journal}
  {\bibinfo  {journal} {{Nat. Nanotech.}}\ }\textbf {\bibinfo {volume} {13}},\
  \bibinfo {pages} {362} (\bibinfo {year} {2018})}\BibitemShut {NoStop}%
\bibitem [{\citenamefont {Grigorev}\ \emph {et~al.}(2021)\citenamefont
  {Grigorev}, \citenamefont {Filianina}, \citenamefont {Bodnar}, \citenamefont
  {Sobolev}, \citenamefont {Bhattacharjee}, \citenamefont {Bommanaboyena},
  \citenamefont {Lytvynenko}, \citenamefont {Skourski}, \citenamefont {Fuchs},
  \citenamefont {Kl\"aui}, \citenamefont {Jourdan},\ and\ \citenamefont
  {Demsar}}]{Grigorev2021}%
  \BibitemOpen
  \bibfield  {author} {\bibinfo {author} {\bibfnamefont {V.}~\bibnamefont
  {Grigorev}}, \bibinfo {author} {\bibfnamefont {M.}~\bibnamefont {Filianina}},
  \bibinfo {author} {\bibfnamefont {S.~Y.}\ \bibnamefont {Bodnar}}, \bibinfo
  {author} {\bibfnamefont {S.}~\bibnamefont {Sobolev}}, \bibinfo {author}
  {\bibfnamefont {N.}~\bibnamefont {Bhattacharjee}}, \bibinfo {author}
  {\bibfnamefont {S.}~\bibnamefont {Bommanaboyena}}, \bibinfo {author}
  {\bibfnamefont {Y.}~\bibnamefont {Lytvynenko}}, \bibinfo {author}
  {\bibfnamefont {Y.}~\bibnamefont {Skourski}}, \bibinfo {author}
  {\bibfnamefont {D.}~\bibnamefont {Fuchs}}, \bibinfo {author} {\bibfnamefont
  {M.}~\bibnamefont {Kl\"aui}}, \bibinfo {author} {\bibfnamefont
  {M.}~\bibnamefont {Jourdan}}, \ and\ \bibinfo {author} {\bibfnamefont
  {J.}~\bibnamefont {Demsar}},\ }\href {\doibase
  10.1103/PhysRevApplied.16.014037} {\bibfield  {journal} {\bibinfo  {journal}
  {Phys. Rev. Applied}\ }\textbf {\bibinfo {volume} {16}},\ \bibinfo {pages}
  {014037} (\bibinfo {year} {2021})}\BibitemShut {NoStop}%
\bibitem [{\citenamefont {H{\"u}fner}(2003)}]{Hufner2003}%
  \BibitemOpen
  \bibfield  {author} {\bibinfo {author} {\bibfnamefont {S.}~\bibnamefont
  {H{\"u}fner}},\ }\href@noop {} {\emph {\bibinfo {title} {{Photoelectron
  Spectroscopy - Principles and Applications}}}}\ (\bibinfo  {publisher}
  {Springer},\ \bibinfo {address} {Berlin},\ \bibinfo {year}
  {2003})\BibitemShut {NoStop}%
\bibitem [{\citenamefont {Reinert}\ and\ \citenamefont
  {H{\"u}fner}(2005)}]{Reinert2005}%
  \BibitemOpen
  \bibfield  {author} {\bibinfo {author} {\bibfnamefont {F.}~\bibnamefont
  {Reinert}}\ and\ \bibinfo {author} {\bibfnamefont {S.}~\bibnamefont
  {H{\"u}fner}},\ }\href@noop {} {\bibfield  {journal} {\bibinfo  {journal}
  {New J. Phys.}\ }\textbf {\bibinfo {volume} {7}},\ \bibinfo {pages} {97}
  (\bibinfo {year} {2005})}\BibitemShut {NoStop}%
\bibitem [{\citenamefont {Fiebig}\ \emph {et~al.}(1994)\citenamefont {Fiebig},
  \citenamefont {Frohlich}, \citenamefont {Krichevtsov},\ and\ \citenamefont
  {Pisarev}}]{Fiebig1994}%
  \BibitemOpen
  \bibfield  {author} {\bibinfo {author} {\bibfnamefont {M.}~\bibnamefont
  {Fiebig}}, \bibinfo {author} {\bibfnamefont {D.}~\bibnamefont {Frohlich}},
  \bibinfo {author} {\bibfnamefont {B.~B.}\ \bibnamefont {Krichevtsov}}, \ and\
  \bibinfo {author} {\bibfnamefont {R.~V.}\ \bibnamefont {Pisarev}},\
  }\href@noop {} {\bibfield  {journal} {\bibinfo  {journal} {Phys. Rev. Lett.}\
  }\textbf {\bibinfo {volume} {73}},\ \bibinfo {pages} {2127} (\bibinfo {year}
  {1994})}\BibitemShut {NoStop}%
\bibitem [{\citenamefont {Sch{\"{u}}tte-Engel}\ \emph
  {et~al.}(2021)\citenamefont {Sch{\"{u}}tte-Engel}, \citenamefont {Marsh},
  \citenamefont {Millar}, \citenamefont {Sekine}, \citenamefont {Chadha-Day},
  \citenamefont {Hoof}, \citenamefont {Ali}, \citenamefont {Fong},
  \citenamefont {Hardy},\ and\ \citenamefont {{\v{S}}mejkal}}]{Engel2021}%
  \BibitemOpen
  \bibfield  {author} {\bibinfo {author} {\bibfnamefont {J.}~\bibnamefont
  {Sch{\"{u}}tte-Engel}}, \bibinfo {author} {\bibfnamefont {D.~J.}\
  \bibnamefont {Marsh}}, \bibinfo {author} {\bibfnamefont {A.~J.}\ \bibnamefont
  {Millar}}, \bibinfo {author} {\bibfnamefont {A.}~\bibnamefont {Sekine}},
  \bibinfo {author} {\bibfnamefont {F.}~\bibnamefont {Chadha-Day}}, \bibinfo
  {author} {\bibfnamefont {S.}~\bibnamefont {Hoof}}, \bibinfo {author}
  {\bibfnamefont {M.~N.}\ \bibnamefont {Ali}}, \bibinfo {author} {\bibfnamefont
  {K.~C.}\ \bibnamefont {Fong}}, \bibinfo {author} {\bibfnamefont
  {E.}~\bibnamefont {Hardy}}, \ and\ \bibinfo {author} {\bibfnamefont
  {L.}~\bibnamefont {{\v{S}}mejkal}},\ }\href {\doibase
  10.1088/1475-7516/2021/08/066} {\bibfield  {journal} {\bibinfo  {journal}
  {Journal of Cosmology and Astroparticle Physics}\ }\textbf {\bibinfo {volume}
  {2021}},\ \bibinfo {pages} {066} (\bibinfo {year} {2021})}\BibitemShut
  {NoStop}%
\bibitem [{\citenamefont {Sch{\"o}nhense}\ \emph {et~al.}(2020)\citenamefont
  {Sch{\"o}nhense}, \citenamefont {Babenkov}, \citenamefont {Vasilyev},
  \citenamefont {Elmers},\ and\ \citenamefont {Medjanik}}]{Schonhense2020}%
  \BibitemOpen
  \bibfield  {author} {\bibinfo {author} {\bibfnamefont {G.}~\bibnamefont
  {Sch{\"o}nhense}}, \bibinfo {author} {\bibfnamefont {S.}~\bibnamefont
  {Babenkov}}, \bibinfo {author} {\bibfnamefont {D.}~\bibnamefont {Vasilyev}},
  \bibinfo {author} {\bibfnamefont {H.-J.}\ \bibnamefont {Elmers}}, \ and\
  \bibinfo {author} {\bibfnamefont {K.}~\bibnamefont {Medjanik}},\ }\href
  {\doibase 10.1063/5.0024074} {\bibfield  {journal} {\bibinfo  {journal} {Rev.
  Sci. Instr.}\ }\textbf {\bibinfo {volume} {91}},\ \bibinfo {pages} {123110}
  (\bibinfo {year} {2020})},\ \Eprint
  {http://arxiv.org/abs/https://doi.org/10.1063/5.0024074}
  {https://doi.org/10.1063/5.0024074} \BibitemShut {NoStop}%
\bibitem [{\citenamefont {Dewhurst}()}]{elk}%
  \BibitemOpen
  \bibfield  {author} {\bibinfo {author} {\bibfnamefont {K.}~\bibnamefont
  {Dewhurst}},\ }\href {http://elk.sourceforge.net} {\enquote {\bibinfo {title}
  {{Elk code}},}\ }\BibitemShut {NoStop}%
\bibitem [{\citenamefont {Kawamura}(2019)}]{Kawamura2019}%
  \BibitemOpen
  \bibfield  {author} {\bibinfo {author} {\bibfnamefont {M.}~\bibnamefont
  {Kawamura}},\ }\href {\doibase 10.1016/J.CPC.2019.01.017} {\bibfield
  {journal} {\bibinfo  {journal} {Computer Physics Communications}\ }\textbf
  {\bibinfo {volume} {239}},\ \bibinfo {pages} {197} (\bibinfo {year}
  {2019})}\BibitemShut {NoStop}%
\end{thebibliography}
%

\section{Supplementary Material and Methods}

\subsection{Experimental Details}
{ Samples were transported from the deposition chamber to the photoemission experiment 
using an ultra-high vacuum suitcase.
For ARPES measurements, photoelectrons were excited by a He discharge lamp (21.2~eV) and by a pulsed laser (6.4~eV, 80~MHz repetition rate, APE).
The incidence angle of the photon beam is 22$^0$ with respect to the sample surface 
along the $x$-axis.
The samples have been aligned such that the $x$ and $y$ directions correspond to 
the magnetic $\langle 110 \rangle$ easy axes, respectively. 
Photoemission experiments at 21.2~eV have been performed using the single-hemisphere momentum microscope described in Ref.~\cite{Schonhense2020} with the energy resolution set to 50~meV and
laser ARPES experiments using a ToFMM (Surface Concept GmbH) with the resolution set to 40~meV.
For the latter experiment, a field aperture inserted at the position of a Gaussian image allows to restrict the region of interest to a circular area with 0.9~$\mu$m diameter,
while the downstream electron optics can be set from Gaussian to Fourier imaging
for momentum-mapping.}

\subsection{Theoretical Methods}
For the equilibrium density functional theory calculations and
symmetry analysis, we used the FLAPW code ELK {\cite{elk}}.
We used the BCT unit cell and a k-point mesh $10 \times 10 \times 10$. We plot
the Fermi surfaces with the program Fermisurfer {\cite{Kawamura2019}}.
More details are given in Refs.~\onlinecite{Bodnar2018,Elmers2020}.
\end{document}